\documentclass[epsf,useAMS,usenatbib,usegraphicx]{mn2e} 
\usepackage{epsf} 
\usepackage[usenames]{color} 
 \newcommand{\changea}{}

\title[Radial velocity study of HD\,3980]{A search for rapid pulsations 
in the magnetic cool chemically peculiar star 
HD\,3980\thanks{Based on observations 
collected at the European Southern Observatory, Paranal, Chile, as 
part of programme 077.D-0150(A) and part of programmes 074.D-0392(A) 
\changea{and 076.D-0535(A) in the ESO Archive.}}}

\author[Elkin et al.] 
{V.G.~Elkin$^{1}$, D.W.~Kurtz$^{1}$, L.M.~Freyhammer$^{1}$, S.~Hubrig$^{2}$ 
and G.~Mathys$^{2}$ \\ 
$^{1}$Centre for Astrophysics, University of Central Lancashire, 
Preston PR1 2HE, UK\\ 
$^{2}$European Southern Observatory, Casilla 19001, Santiago 19, 
Chile} 
\begin{document} 

\maketitle 

\begin{abstract} 
The Ap star HD\,3980 appears to be a \changea{promising roAp candidate}  based on 
its fundamental parameters, leading us to search for rapid pulsations with the VLT 
UV-Visual Echelle Spectrograph (UVES). A precise {\it Hipparcos} parallax and 
estimated temperature of 8100\,K place HD\,3980 in the middle of the theoretical 
instability strip for rapidly oscillating Ap stars, about halfway through its main 
sequence evolution stage. The star has a strong, variable magnetic field, as is 
typical of the cool magnetic Ap stars.  Dipole model parameters were determined 
from \changea{VLT observations using FORS1}. From Doppler shift measurements for 
individual spectral lines of rare earth elements and the H$\alpha$ line core, we 
find no pulsations above $20 - 30$\,m\,s$^{-1}$. This result is corroborated by 
inspection of lines of several other chemical elements, as well as with cross-
correlation
 for long spectral regions with the average spectrum as a template. 
Abundances of chemical elements were determined and show larger than solar 
abundances of rare earth elements. Further, ionisation disequilibria for the first 
two ionised states of Nd and Pr are detected. We also find that the star has a 
strong overabundance of manganese, which is typical for much hotter HgMn and other 
Bp stars. Line profile variability with the rotation period was detected for the 
majority of chemical species. 

\end{abstract} 


\section{Introduction} 

After the discovery of low amplitude pulsation for the bright and well-studied 
cool, magnetic chemically peculiar A (Ap) star $\beta$\,CrB 
\citep{hatzesetal04,kurtzetal07}, the question arose whether all cool Ap stars 
(with effective temperatures below $T_{\rm eff}\sim 8200$\,K) are rapid 
oscillators. This still unanswered question could be a pivotal point for 
theoretical modelling and understanding of the driving mechanism in rapidly 
oscillating (roAp) stars. Several searches for rapid oscillations in Ap stars have 
been made. \citet{elkinetal08a} found that 24 known roAp stars that exhibit 
pulsations photometrically also show rapid radial velocity variations for the 
corresponding pulsation periods. However, several Ap stars with photometric 
indices typical for known roAp stars were tested photometrically for pulsations by 
\citet{martinezetal94} and found to be stable to high precision; these stars are 
called non-oscillating Ap stars, or noAp stars. 

About a decade later, \citet{elkinetal05a} used fast UVES spectroscopy to discover 
low amplitude radial velocity pulsation for one of these stars, the former noAp 
star HD\,116114. \citet{lorenzetal05} reported a new photometric null result for 
the star, but noted a low amplitude pulsation peak in the amplitude spectrum of 
their data slightly above the noise level for the same frequency. Spectroscopy 
therefore is superior to photometry as a tool for detection of rapid low amplitude 
pulsation, and we expect that more Ap stars previously identified as non-
oscillating (noAp) may exhibit rapid radial velocity variations. One such 
promising example is HD\,965, which has physical parameters corresponding to those 
of known roAp stars. The fast photometry by \citet{kurtzetal03} and high time 
resolution spectroscopy by \citet{elkinetal05b}, however, found no rapid 
variability. Both of these studies mentioned the known roAp problem of amplitude 
modulation with the rotation phase as a possible explanation for the null result 
and proposed to re-observe the star at a different rotation phase when one of the 
magnetic poles is at a more favourable aspect. 

As seen in, e.g., the astrometric HR-diagram by \citeauthor{hubrigetal05}\, 
(\citeyear{hubrigetal05}, their figure 2) for roAp and noAp stars, the apparent 
noAp stars occupy essentially the same regions as the roAp stars. However, the 
noAp stars appear to be systematically more evolved than the roAp stars 
\citep{northetal97,handleretal99,hubrigetal00}. Nevertheless, the theoretical roAp 
instability strip \citep{cunha02} also predicts rapid pulsations for the more 
evolved Ap stars (near the terminal age the main sequence). HD\,116114 
\citep{elkinetal05b} was indeed detected in this region of the HR-diagram, 
oscillating with the predicted frequency 0.79\,mHz (the lowest frequency known for 
the roAp stars). Still the only case of a luminous roAp star, HD\,116114 shows 
extremely low radial velocity pulsation amplitude and only for small number of 
chemical elements such as europium and lanthanum. 

Using the same instrument (UVES) and procedure as \citet{elkinetal05b}, 
\citet{freyhammeretal08} searched for rapid pulsations among a group of nine 
evolved Ap stars inside the roAp instability region in the HR-diagram. 
Surprisingly they did not detect any radial velocity pulsation for these stars, 
but showed that only $3-5$ of the stars may have magnetic field strengths 
considerably in \changea{excess of 2\,kG}. More evolved stars are theoretically expected 
to require relatively strong magnetic fields to suppress local surface convection 
and facilitate observable amplitudes of rapid pulsations (see \citealt{cunha02}). 

\begin{figure} 
\begin{center} 
\hfil \epsfxsize 8.5cm\epsfbox{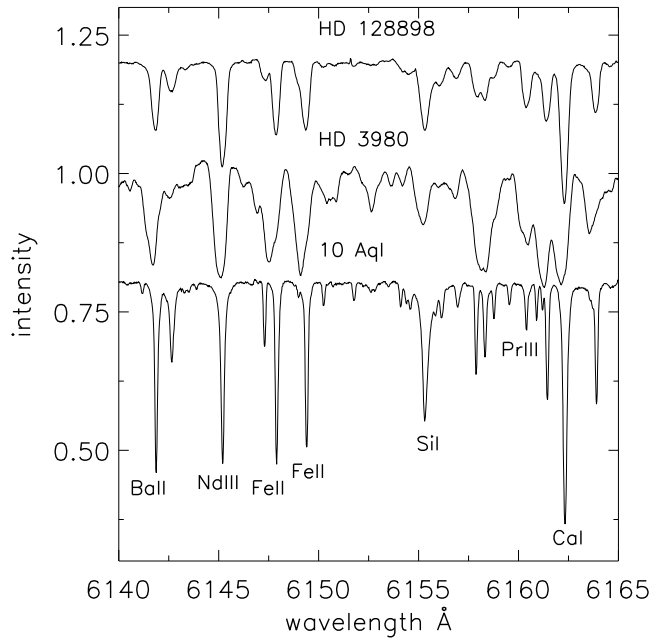} 
\caption{\label{Fig1}A small spectral region of HD\,3980, compared with those of 
the two roAp stars, HD\,128898 ($\alpha$\,Cir) and HD\,176232 (10\,Aql). The 
spectra of the latter two stars are shifted up and down, respectively, in 
intensity by 0.2. Selected atomic lines are indicated. HD\,3980 shows strong lines 
for the rare earth elements Nd and Pr, a common characteristic of known roAp 
stars. } 
\end{center} 
\end{figure}

The bright Ap star HD\,3980 ($\xi$\,Phe, $V=5.719$ mag) was proposed by one of us 
(Hubrig) to be a good roAp candidate because of its many properties in common with 
those of known roAp stars, such as a magnetic field and peculiar abundances of 
rare earth elements, as we show in Fig.\,\ref{Fig1}. A photometric search for 
rapid pulsations had already been performed for HD\,3980 by \citet{martinezetal94} 
who failed to detect any pulsational variability. Based on our success with high 
time resolution spectroscopy (e.g. \citealt{kurtzetal07,mathysetal07}), we then 
decided to use fast spectroscopy to test this promising roAp candidate for rapid 
pulsations. The following sections discuss the collection of data, their analysis 
and the null result for rapid oscillations. 

\section{Observations and data reduction} 

HD\,3980 was observed twice in 2006 July with UVES on the ESO Very Large Telescope 
(VLT) to search for rapid pulsations at two rotation phases roughly coinciding 
with the predicted times of magnetic maximum and minimum (see 
Sect.\,\ref{sec:stelpar}), when pulsation amplitudes are expected to be highest. 
On the first observing night, 2006 July 23, 104 spectra were obtained in 2.0\,hr, 
and on 2006 July 25, another 55 spectra were collected in 1.0\,hr. The seeing was 
relatively poor, especially on the first night when it varied from 
$1\rlap.^{\prime\prime}5$ to 3$^{\prime\prime}$. However, the star is bright and 
we used an image slicer (IS$\#3$) to optimally utilise the observing conditions at 
the instrument's maximum resolution with a 0\farcs3 slit. Exposure times of 40\,s 
were then used, which together with readout and overhead times of $\sim28$\,s 
provided a time resolution of 68\,s. We used the RED (600\,nm) setting which 
covers the wavelength region $\lambda\lambda\,4970 - 7010$\,\AA, with a gap in the 
region $\lambda\lambda\,5963 - 6032$\,\AA\ caused by the space between the two CCD 
mosaic halves. The average spectral resolution is about $R = 110\,000$. The camera 
uses two 2K $\times$ 4K CCDs with 15\,$\mu$m pixels. Raw CCD frames were processed 
using the UVES pipeline to extract and merge the echelle orders to 1D spectra that 
were normalised to the continuum. The average signal-to-noise ratio in the 
continuum, estimated from 1D spectra, is about 250. 

\section{The stellar parameters} 
\label{sec:stelpar} 

The rotation period was determined from HD\,3980's double wave light curve by 
\citet{maitzenetal80}, who provided the elements (for the primary minimum): 
JD(Prim. Min. $v)=244\,2314\fd48 + (3\fd9516 \pm 0\fd0003)$\,E. These authors also 
obtained 8 longitudinal magnetic field measurements which show large scatter when 
phased with the rotation period. For the rotation phases $0.30-0.32$, 3 
measurements give $\left<B_{\rm l}\right>=-780\pm700$\,G, while 4 measurements at 
phases $0.78-0.83$ give $860\pm996$\,G. A single point measured near phase zero, 
with a conservative error estimated from the other measurements, gives 
$1670\pm1000$\,G. These 8 measurements are insufficient to \changea{fit the 
magnetic field curve properly}. 

\begin{figure} 
\begin{center} 
\hfill\includegraphics[height=8.3cm, angle=90]{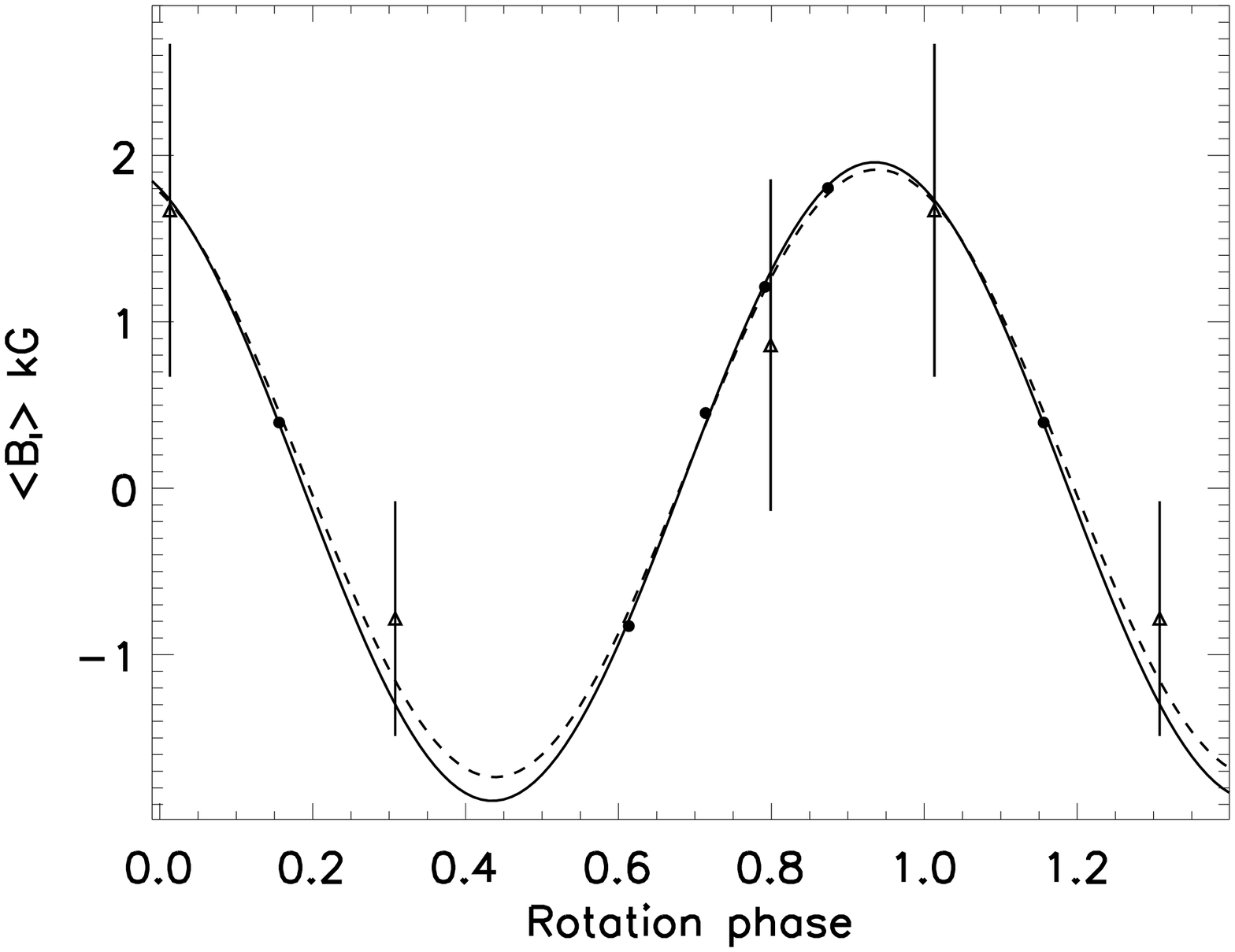} 
\caption{\label{Fig2}
 The longitudinal field observations available for HD\,3980.
The five filled circles are measurements from \citet{hubrigetal06}
and two recent points observed on JD245\,4432.51 and JD245\,4433.54
(Hubrig, unpublished). The full line shows the least squares
fit of a sine function to these 5 points, using phases based
on the light curve elements by \citet{maitzenetal80}. For
comparison, 3 measurements (triangles) by \citet{maitzenetal80}
are also shown with standard deviation error bars and a weighted fit (dashed
line) to all 8 points. Good agreement with the recent and
old data is seen, in spite of the long time baseline.}
\end{center} 
\end{figure}

 \citet{hubrigetal06} obtained 3 precise measurements of the longitudinal magnetic 
field: $\left<B_{\rm l}\right>=1210\pm32$\,G,  $395\pm26$\,G  and $452\pm15$\,G. 
Two more observations were \changea{obtained by Hubrig, using the same method:
$\left<B_{\rm l}\right>=-828\pm18$\,G (JD245\,4432.51) and $1804\pm15$\,G
(JD245\,4433.54)}. 
These measurements are, together with averaged data from \citet{maitzenetal80}, 
shown in Fig.\,\ref{Fig2} with error bars and phased with the rotation period. A 
least squares sine curve fit of \changea{the five Hubrig measurements} provides a mean 
field of 
$\langle B_{\rm l} \rangle=40\pm18$\,G and amplitude of $A_{B_{\rm 
l}}=1918\pm29$\,G. By intention, our two observing nights occurred near the times 
of magnetic extrema at rotation phases 0.94 and 0.46, respectively, which are 
close 
to the positive extremum and the negative extremum of the magnetic field curve. 

\citet{hubrigetal07} determined stellar parameters: $T_{\rm eff} = 8240\pm310$\,K, 
$\log g = 4.05\pm0.09$, $\log L/{\rm L}_\odot=1.296\pm0.052$ and a projected 
rotational velocity $v \sin i = 15 $\,km\,s$^{-1}$. For the 3.95\,d rotation 
period and the longitudinal field measurements, they estimated the magnetic field 
geometry parameters: Mean longitudinal field $\overline{\left< B_l 
\right>}=1200$\,G, inclination angle between rotation axis and line-of-sight $i = 
32^\circ$ and angle between rotation axis and magnetic dipole axis $\beta = 
88^\circ$. These values, however, are highly uncertain.

Our fit to the magnetic measurements shown in Fig.~\ref{Fig2} can also be used to 
constrain the inclinations of the magnetic and rotation axes. It is easy to show 
for a centred dipolar magnetic field that 
\begin{equation}
B_{\rm l} \propto B_{\rm p} \cos \alpha
\end{equation}
\noindent and
\begin{equation}
\cos \alpha = \cos i \cos \beta + \sin i \sin \beta \cos \Omega t
\end{equation}
\noindent where $\alpha$ in the angle between the magnetic pole and the line-of-sight, 
$i$ is the rotational inclination, $\beta$ is the angle between the 
rotation axis and the magnetic axis, $\Omega$ is the rotation frequency, $B_{\rm 
l}$ is the longitudinal magnetic field strength and $B_{\rm p}$ is the polar field 
strength. 

It is obvious from Eq.~2 that the mean magnetic field strength
\begin{equation}
\langle B_{\rm l} \rangle \propto \cos i \cos \beta
\end{equation}
\noindent and the amplitude of the magnetic field variations
\begin{equation}
A_{B_{\rm l}} \propto \sin i \sin \beta\;.
\end{equation}
\noindent Thus we get
\begin{equation}
\tan i \tan \beta = A_{B_{\rm l}}/\langle B_{\rm l} \rangle\;,  
\end{equation}
\noindent from which $\beta$ can be constrained when $i$ is known, using
the values for $A_{B_{\rm l}}$ and $\langle B_{\rm l} \rangle$ from our fit to the 
magnetic data shown in Fig.~\ref{Fig2}. We derive in Sect.~6 below from the {\it 
Hipparcos} parallax and our estimate of $T_{\rm  eff}$ that $R = 
2.19^{+0.19}_{-0.16}$\,R$_{\odot}$, which coupled to our measurement of $v \sin i = 
21.0\pm3.0$\,km\,s$^{-1}$  gives a weak constraint on the rotational inclination 
of $i = 49\degr^{+19}_{-12}$. Eq.~5 can then be used to estimate the magnetic 
obliquity to be \changea{$\beta = 88\fdg6^{+0.8}_{-4.0}$ (accounting for all 1-$\sigma$ 
errors, while it becomes undetermined for 2-$\sigma$ errors). Several 
more precise magnetic measurements at different rotation phases} are still 
needed to fill \changea{the gaps in the magnetic} field curve. The important conclusion here is that 
the obliquity of the magnetic field must be near to $90^\circ$ to account for the 
mean magnetic field strength being close to zero. 

\changea{The stellar parameters $T_{\rm eff}$ and $\log g$ were estimated} from 
Str\"omgren photometry by \citet{martinez93} using the \citet{moonetal85} 
calibrations. Then a small grid of synthetic H$\alpha$ line profiles was compared 
to the observed spectra. The model spectra were produced using Kurucz model 
atmospheres \citep{kurucz79}, models from the NEMO database \citep{heiteretal02} 
and calculations with the {\small SYNTH} program by \citet{piskunov92}. It is 
difficult to locate accurately the continuum in the broad H$\alpha$ region. The 
resulting normalised profiles are slightly different for the two observing sets. 
This may partly be due \changea{to our} continuum placement, but \changea{the 
H$\alpha$ profile may itself be variable.} An example of this was found in the hotter 
peculiar star 36\,Lyn by \citet{takadaeta89}.  Variation in the H$\alpha$ profile 
is consistent with \changea{small variations in the effective temperature of the star
as it is rotates}.  For the average spectrum of the first observing night, 
we obtained a good fit for $T_{\rm eff} = 8000$\,K and $\log g = 4.0$, whereas for 
the second night $T_{\rm eff} = 8200$\,K was required. 

We performed an abundance analysis of the first set of spectra of the star, for 
which $T_{\rm eff} = 8000 \pm 200$\,K and $\log g = 4.0 \pm 0.2$. A Kurucz model 
atmosphere with $T_{\rm eff} = 8000$\,K and $\log g = 4.0$ and with enhanced 
(above solar) metallicity ([M/H]$ =+0.5$) was used for calculating model spectra 
with {\small SYNTH}. Spectral line lists were taken from the Vienna Atomic Line 
Database (VALD, \citealt{kupkaetal99}) and the DREAM database 
\citep{biemontetal99}. We determined $v \sin i = 21.0\pm3.0$\,km\,s$^{-1}$ from 
several symmetric lines. This value is slightly higher than the $v \sin i = 15 
$\,km\,s$^{-1}$ found by \citet{hubrigetal07}. That could be explained by, e.g., a 
variable magnetic field strength which contributes to the broadening of lines with 
large Land\'e \changea{factors. A more probable explanation is line profile variations}
due to a spotted abundance distribution. Some spectral lines are indeed particularly narrow 
and give a lower value of $v \sin i$. \changea{We consider this to be an effect of these
elements being concentrated in spots on the stellar surface such that their
spectral lines do not reflect the full rotational broadening.} An example is 
Li\,\textsc{i} for which spotted distributions have been demonstrated for roAp 
stars by \citet{faraggianaetal96} and \citet{polosukhinaetal00}. For the strong, 
isolated and symmetric but non-resolved doublet Li\,\textsc{i} 6707.76 and 
6707.91\,\AA\ we \changea{obtained} $v \sin i = 16.5 $\,km\,s$^{-1}$. 

\section{Search for rapid radial velocity variations} 

We searched for pulsations as periodic Doppler shifts in two ways: first by using 
cross-correlations of long stretches of spectral regions, then by measuring 
centre-of-gravity shifts of individual lines in the spectra. For frequency 
analyses, we used {\small MIDAS}'s TSA (Time Series Analysis) context, a discrete 
Fourier transform programme by \citet{kurtz85} and the {\mbox{\small PERIOD04}} 
\citep{lenzetal05} programme. 

\subsection{Cross-correlation radial velocity analyses} 

The cross-correlation method, using large spectral regions, 
\changea{is often useful} for detecting pulsation in roAp star candidates 
and for finding additional 
frequencies in known roAp stars (see, e.g., \citealt{matthewsetal88}; 
\citealt{balonaetal02}; \citealt{hatzesetal04}).  
The cross-correlation amplitudes from correlation of 
long spectral regions are, though, not directly comparable to those derived from 
line profile measurements. This is mainly due to \changea{different pulsation 
amplitudes and phases of different ions}
 in the stratified roAp atmospheres, where low 
amplitude elements such as Fe dilute the `integrated' Doppler shifts. 
\changea{An example is the roAp star 10\,Aql for which \citet{elkinetal08b} 
found} a pulsation amplitude of $6 \pm 1$\,m\,s$^{-1}$ with cross-correlation for the 
spectral region $5150-5450$\,\AA, while individual lines of rare earth elements 
showed amplitudes in excess of 500\,m\,s$^{-1}$. 
10\,Aql represents a case of a roAp star for which the lines of 
rare earth ions are weak. Another example is HD\,154708 which has a 
very strong magnetic field.  
From cross-correlation measurements of HD\,154708
  in the spectral range  $5150-5800$\,\AA\  \citet{freyhammeretal08} 
detected pulsation  with an amplitude of
  10\,m\,s$^{-1}$, while \citet{kurtzetal06a} 
 obtained amplitudes around 60\,m\,s$^{-1}$ for some individual rare earth lines.

Cross-correlations were performed with our HD\,3980 spectra, using the average 
spectrum as template. For the line-rich spectral range $5000-5960$\,\AA, we 
searched the regions $5150-5400$\,\AA\ and $5400-5700$\,\AA, but in both cases no 
significant signal was detected above a level of 10\,m\,s$^{-1}$ with 
$\sigma=3$\,m\,s$^{-1}$ (see Fig.\,\ref{Fig3}, top panel) in the frequency domain 
of known roAp stars. Only low frequency peaks below 0.4\,mHz are seen. However, we 
disregard these periodicities as they are also seen in the radial velocity 
measurements of telluric lines, hence are instrumental in origin. A spectral 
region from $6863 - 6938$\,\AA\ with abundant telluric lines was used to check the 
instrumental stability and identify non-stellar periodicities. The results from 
telluric lines only show low frequency noise (due to instrumental drifts or 
meteorological changes). The telluric region otherwise shows stability at the 
level below 10\,m\,s$^{-1}$ with $\sigma=3$\,m\,s$^{-1}$.  The spectral region 
longwards of 6000\,\AA\ has a lower line density and a higher scatter, but 
confirms these null results. 

\subsection{Line profile radial velocity analyses} 

\begin{figure} 
\begin{center} 
\hfil \epsfxsize 8.5cm\epsfbox{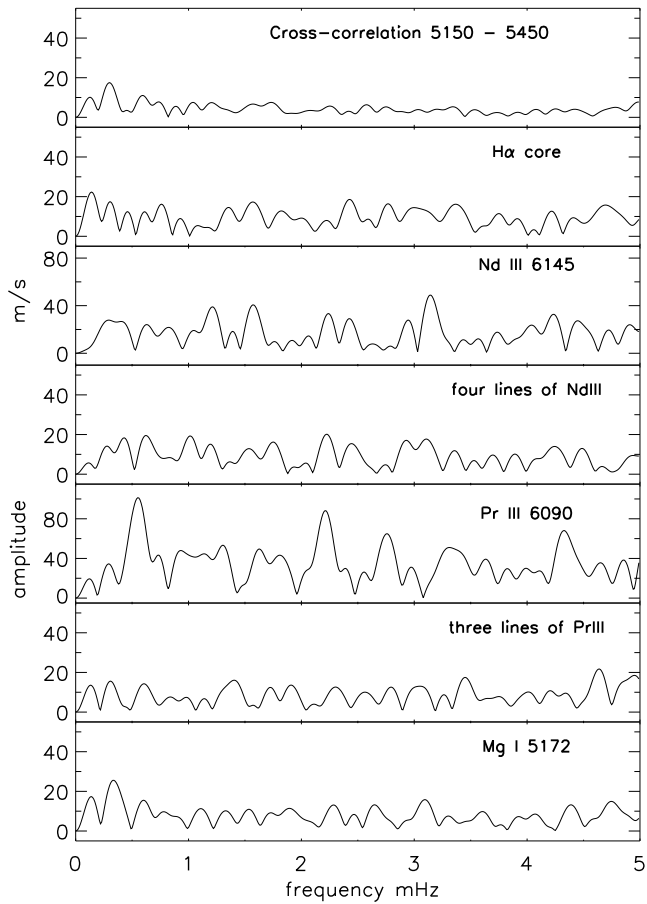} 
\caption{\label{Fig3} Amplitude spectra for spectral lines of HD\,3980. there are 
no significant peaks. } 
\end{center} 
\end{figure}

In roAp stars, lines of rare earth elements typically show the largest Doppler 
shift pulsation amplitudes. Amplitudes vary for different elements and range from 
a few dozen metres per second up to a few kilometres per second for various roAp 
stars \citep{elkinetal08a}. Also the narrow line core of the H$\alpha$ profile 
shows rapid pulsations in roAp stars \citep{kurtzetal06a,elkinetal08a}. We 
therefore searched for pulsations in HD\,3980 using \changea{the centre of gravity
method for similar} lines and show 
amplitude spectra of selected lines in Fig.\,\ref{Fig3}. Although we concentrate 
on analyses of lines of the rare earths, other chemical elements were also tested. 
All radial velocity curves subjected to period searches were \changea{de-trended 
beforehand} with linear least square fitting, which also eliminated the barycentric 
velocity correction which is approximately linear during the short duration of our 
two runs. The H$\alpha$ core was stable above the 20\,m\,s$^{-1}$ with a noise 
level of 7\,m\,s$^{-1}$. 

Of the rare earth elements, we examined lines of Ce\,\textsc{ii}, 
Pr\,\textsc{iii}, Nd\,\textsc{iii}, Nd\,\textsc{ii}, Eu\,\textsc{ii} and 
Gd\,\textsc{ii}, but no pulsations were detected from any of these lines above 
typical upper limits of 30\,m\,s$^{-1}$ (with noise levels varying from 10 to 
30\,m\,s$^{-1}$ for the majority of good lines). Combination of $3-4$ lines 
reduced the noise level of individual lines,  but also did not show any reliable 
pulsation signal. Due to the rotational broadening and many asymmetric line 
profiles, the number of suitable strong, symmetric lines was rather limited. Of 
the non-rare-earth element lines analysed, including the strong sodium D and 
Mg\,\textsc{i} lines and several Ca, Sc, Ti, Cr, Fe and Ba lines, no pulsations 
were detected above $15 - 30$\,m\,s$^{-1}$. In general, some lines show intriguing 
peaks in the amplitude spectra for the frequency range typical for roAp stars. But 
as these peaks could not be confirmed by other spectral lines of the same element 
or by other rare earth lines, they were rejected as indications of pulsation in 
HD\,3980. Examples of such peaks are seen in Fig.\,\ref{Fig3} for the asymmetric 
and blended line of Pr\,\textsc{iii} 6090\,\AA\ which shows two peaks with 
frequencies 0.546\,mHz and 2.752\,mHz. Combining radial velocity measurements of 
three other Pr\,\textsc{iii} lines (seen in the panel below that for 
Pr\,\textsc{iii} 6090\,\AA) shows no significant peaks for these two frequencies 
nor for any others. 

Among the Ap stars, HD\,3980 has strong lines of lithium (e.g. equivalent width of 
101\,m\AA\ for Li\,\textsc{i} 6708\,\AA). We therefore also searched the 
Li\,\textsc{i} 6708\,\AA\ doublet for pulsation. This line shows strong 
variability with rotation phase, but is stable to rapid pulsations above 
30\,m\,s$^{-1}$ with a noise level of 11\,m\,s$^{-1}$. 

\subsection{Linear trends in line profile radial velocity shifts} 

Line shapes of Nd, Eu and Li are observed to change strongly with rotation phase 
in HD\,3980, such as the example in Fig.\,\ref{Fig4} with Nd\,\textsc{iii} 
6145\,\AA. This is a clear indication of an inhomogeneous surface distributions of 
elements that appear to be located in patches. The line analysis is complicated by 
the resulting line asymmetries, as well as by many lines that are blends because 
of HD\,3980's relatively high rotation velocity and strong lines of peculiar 
elements.

A large number of spectral lines show quasi-linear drifts in radial velocities 
which appear as low frequency peaks in their amplitude spectra. Different lines 
may show different drifts over the interval of $1-2$\,hr. We have noticed the same 
effect for several other Ap and roAp stars (see, e.g., 
\citealt{freyhammeretal08}).  For HD\,99563, \citet{elkinetal05c} detected and 
discussed even larger trends for particular lines. For the roAp star $\beta$\,CrB, 
\citet{kurtzetal06b} similarly detected non-linear low frequency trends which they 
de-trended from the radial velocity series. For the moderately fast rotating Ap 
\changea{stars, the trends are partly caused by spots} seen at different aspects. For example, 
HD\,99563 has $v \sin i = 30$\,km\,s$^{-1}$ and HD\,3980 has $v \sin i = 
21$\,km\,s$^{-1}$; both these stars show stronger trends than those seen for the 
slower rotator $\beta$\,CrB ($v \sin i = 3.5 $\,km\,s$^{-1}$). We observe that 
these trends also vary from element to element, depending on the particular 
surface distribution. 

For HD\,3980 we de-trended linear drifts prior to the radial velocity analyses. 
However, those trends \changea{clearly contain physical information about} the star's surface 
distribution. The Li\,\textsc{i} 6708\,\AA\ doublet shows one of the strongest 
trends with a linear radial velocity change of about 0.77\,km\,s$^{-1}$\,h$^{-1}$. 
We believe that this is the result of a concentrated strong spot of Li seen at 
\changea{varying aspects} with rotation. Considerable line profile changes \changea{with
the rotation phase are clearly visible in this line during} our 2\,hr run supplemented with 
existing spectra in the ESO Archive at several other rotation phases. \changea{The 
available data only allow a crude tomographic study, but do indicate} different surface 
distributions of various \changea{elements.} 
A full Doppler imaging study \changea{(such as \citealt{lehmannetal07})} 
of this star over its rotation period is needed. 

\subsection{The second observing run} 

The second observing run, obtained about half a rotation period (0.49 phase 
difference) after the first run, collected 55 spectra in 1\,hr. Though only half 
the length of the first run, this set is sufficient for searching for frequencies 
in the known range for roAp stars. As for the previous run, this set also shows 
linear radial velocity drifts of comparable, or slightly higher amplitudes. Again, 
we see drifts for the rare earth element lines with increasing radial velocity, 
while lines of iron peak elements and light elements show no drifts at all, 
consistent with spots for the rare earth elements and a more uniform distribution 
for the lighter elements. 

After de-trending the radial velocity series for individual lines, the frequency 
analysis again showed that there is no pulsation above amplitudes of 
30\,m\,s$^{-1}$ for noise levels of $\sigma=10$\,m\,s$^{-1}$. Cross-correlations of this set 
of spectra with their mean spectrum as template excluded periodic variability 
above 10\,m\,s$^{-1}$ ($\sigma=3$\,m\,s$^{-1}$). 

\begin{figure} 
\begin{center} 
\hfil \epsfxsize 8.5cm\epsfbox{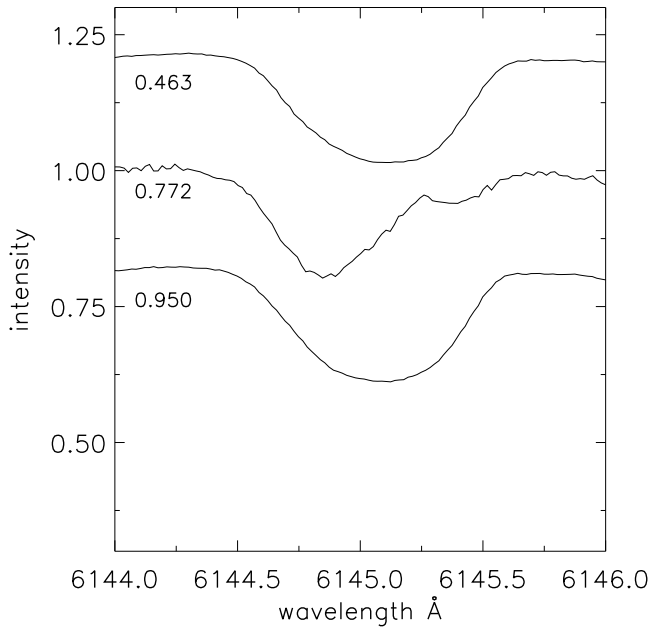} 
\caption{\label{Fig4} Change of the Nd\,\textsc{iii} 6145.07\,\AA\ line profile 
with rotational phase. Our search for pulsation was done for phases corresponding 
to the upper and lower spectra. } 
\end{center} 
\end{figure}

Line profile shapes in the average spectrum are very similar to those of the first 
run, although they cover two opposite sides of the spotted stellar surface at the 
moments of maximum and minimum magnetic field strengths. Spectral lines such as 
Li\,\textsc{ii} and Eu\,\textsc{ii} that also have rather comparable line profile 
shapes, differ strongly from the same lines in spectra from the ESO archive at 
rotation phases 0.2 and 0.8. For example at rotational phase 0.77, the lines of 
Nd\,\textsc{iii} (see, e.g., Nd\,\textsc{iii} 6145\,\AA\ in Fig.\,\ref{Fig4}) have 
double profiles with a sharper, stronger and more blue component. For this phase, 
the data suggest that Nd is concentrated in two different spots, one at each 
magnetic pole. In both our observing runs this line seems more symmetric and wide, 
which may correspond to an extended surface region with overabundance of 
neodymium. 

\section{Chemical abundances} 

The presence of high overabundances of rare earth elements is one of the 
characteristics of known roAp stars. Another important property is that their 
abundances of the first two ionised states of neodymium and praseodymium show more 
than 1\,dex difference, with the doubly ionised ions, which form higher in the 
atmosphere, being the most abundant \citep{ryabchikovaetal04,kurtzetal07}. This 
ionisation disequilibrium anomaly may be explained mostly by concentration of rare 
earth elements in high atmospheric layers (stratification) and partly by non-LTE 
(NLTE) effects \citep{mashonkinaetal05}. NLTE effects may enhance the ionisation 
of Nd\,\textsc{ii} or Pr\,\textsc{ii} and accordingly weaken their absorption 
lines, while strengthening those of the second ionisation state. The 
disequilibrium may not be limited to roAp stars, which are relatively cool Ap 
stars, but may also exist for hotter Ap stars. 

To compare HD\,3980 spectrally with the roAp stars, we measured its chemical 
abundances and tested for ionisation disequilibria for Nd and Pr. Abundances were 
determined by fitting synthetic spectra to the observed average spectra. Model 
spectra were calculated as described in Sect.\,\ref{sec:stelpar}. The resulting 
abundances for the 2\,hr run average spectrum are presented in 
Table\,\ref{Table1}. Solar abundances are from \citet{asplundetal05}. Only lines 
of Mg\,\textsc{i} suggest slightly less than solar abundances. The Fe abundance is 
near solar, or slightly greater. Notably, we detected large overabundances of 
manganese. This element typically shows overabundances for the hotter group of 
peculiar HgMn stars, while for cooler roAp stars, Mn mostly shows abundances 
slightly above solar
 \citep{ryabchikovaetal04}.

\begin{figure*} 
\begin{center} 
\hfil \epsfxsize 14cm\epsfbox{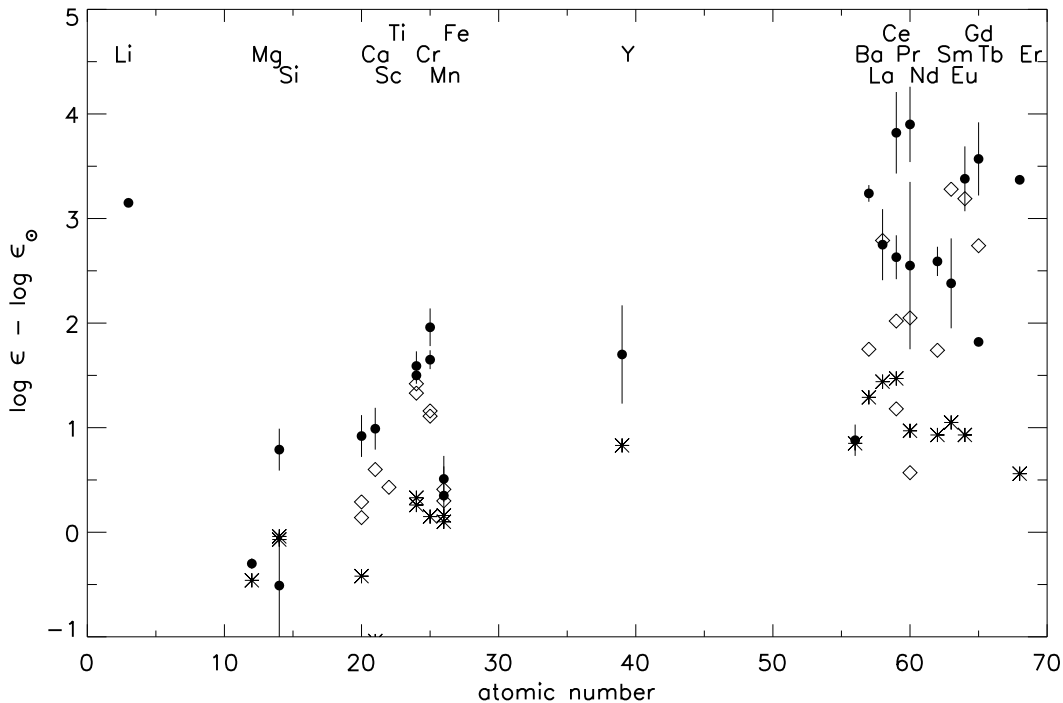} 
\caption{\label{Fig5} Relative abundances ($\log\,\epsilon-\log\,\epsilon_\odot$) 
for HD\,3980 (circles with error bars) compared to those of the roAp star 
$\beta$\,CrB (diamonds) from \citet{kurtzetal07}, and the Am star 32\,Aqr 
(asterisks) by \citet{adelmanetal97} showing the strong overabundances typical of 
the cool Ap stars.}
\end{center} 
\end{figure*}

The rare earth elements and lithium (Fig.\,\ref{Fig5}) show abundances that are 
much greater than solar. We find indications of ionisation disequilibria for both 
neodymium and praseodymium: $\Delta[{\rm Nd}]_{\textsc{iii}  - \textsc{ii}} = 1.4 
\pm 0.3$ and $\Delta[{\rm Pr}]_{\textsc{iii}  - \textsc{ii}} = 1.2 \pm 0.2$
\changea{(where the errors are calculated from the error in the mean abundance for 
each ion's set of lines)}, 
consistent with the spectroscopic signature for most roAp stars discovered by 
\citet{ryabchikovaetal04}. Thus, we find HD\,3980 is spectrally comparable to 
known roAp stars. Three lines of terbium also suggest ionisation disequilibrium 
for this element. 

\begin{table} 
\caption[] {\label{Table1}Chemical abundances for HD\,3980 for selected elements, 
and their corresponding solar abundances \citep{asplundetal05}.  The errors quoted 
are internal standard deviations for the set of lines measured. } 
\begin{center} 
\begin{tabular}{lcrr} 
\hline 
\multicolumn{1}{c}{Ion} 
& \multicolumn{1}{c}{Number} 
& \multicolumn{1}{c}{$\log\,\epsilon$} 
& \multicolumn{1}{c}{$\log\,\epsilon$} \\ 
& \multicolumn{1}{c} {of lines} 
& \multicolumn{1}{c}{HD\,3980} 
& \multicolumn{1}{c}{Sun} \\ 
% & \multicolumn{1}{c}{} \\ 
\hline 
Li\,\textsc{i}    &   1   & 4.20 \phantom{$\pm$ 0.05}&  1.05   $\pm$   0.10 \\ 
Mg\,\textsc{i}    &   3   & 7.23       $\pm$ 0.05  &  7.53     $\pm$ 0.09  \\ 
Si\,\textsc{i}    &   2   & 7.00       $\pm$ 0.50  &  7.51     $\pm$ 0.04  \\ 
Si\,\textsc{ii}   &   2   & 8.30       $\pm$ 0.20  &  7.51     $\pm$ 0.04   \\ 
Ca\,\textsc{i}    &   4   & 7.23       $\pm$ 0.20  &  6.31     $\pm$ 0.04  \\ 
Sc\,\textsc{ii}   &   3   & 4.04       $\pm$ 0.20  &  3.05     $\pm$ 0.08  \\ 
Cr\,\textsc{i}    &   9   & 7.14       $\pm$ 0.08  &  5.64     $\pm$ 0.10  \\ 
Cr\,\textsc{ii}   &   12  & 7.23       $\pm$ 0.14  &  5.64     $\pm$ 0.10  \\ 
Mn\,\textsc{i}    &   7   & 7.04       $\pm$ 0.09  &  5.39     $\pm$ 0.03  \\ 
Mn\,\textsc{ii}   &   6   & 7.35       $\pm$ 0.18  &  5.39     $\pm$ 0.03  \\ 
Fe\,\textsc{i}    &   15  & 7.80       $\pm$ 0.28  &  7.45     $\pm$ 0.05  \\ 
Fe\,\textsc{ii}   &   6   & 7.96       $\pm$ 0.22  &  7.45     $\pm$ 0.05   \\ 
Y \,\textsc{ii}   &   3   & 3.87       $\pm$ 0.47  &  2.17     $\pm$ 0.04  \\ 
Ba\,\textsc{ii}   &   2   & 3.05       $\pm$ 0.15  &  2.17     $\pm$ 0.07  \\ 
La\,\textsc{ii}   &   4   & 4.37       $\pm$ 0.08  &  1.13     $\pm$ 0.05  \\ 
Ce\,\textsc{ii}   &   6   & 4.33       $\pm$ 0.34  &  1.58     $\pm$ 0.09  \\ 
Pr\,\textsc{iii}  &   7   & 4.53       $\pm$ 0.39  &  0.71     $\pm$ 0.08  \\ 
Pr\,\textsc{ii}   &   8   & 3.34       $\pm$ 0.21  &  0.71     $\pm$ 0.08  \\ 
Nd\,\textsc{iii}  &   5   & 5.35       $\pm$ 0.36  &  1.45     $\pm$ 0.05  \\ 
Nd\,\textsc{ii}   &   12  & 4.00       $\pm$ 0.80  &  1.45     $\pm$ 0.05  \\ 
Sm\,\textsc{ii}   &   3   & 3.60       $\pm$ 0.14  &  1.01     $\pm$ 0.06  \\ 
Eu\,\textsc{ii}   &   3   & 2.90       $\pm$ 0.43  &  0.52     $\pm$ 0.06  \\ 
Gd\,\textsc{ii}   &   4   & 4.50       $\pm$ 0.31  &  1.12     $\pm$ 0.04  \\ 
Tb\,\textsc{iii}  &   2   & 3.85       $\pm$ 0.35  &  0.28     $\pm$ 0.30  \\ 
Tb\,\textsc{ii}   &   1   & 2.10 \phantom{$\pm$ 0.05}&  0.28   $\pm$   0.30  \\ 
Er\,\textsc{ii}   &   1   & 4.30 \phantom{$\pm$ 0.05}&  0.93   $\pm$   0.06  \\  
 \hline 
\end{tabular} 
\label{amp} 
\end{center} 
\end{table} 

\section{Discussion} 

The cool Ap star HD\,3980 shows no rapid oscillations in radial velocity above a 
few tens of m\,s$^{-1}$ at two rotational phases separated by half a rotation 
period. We have shown that at least one of these coincides with a rotation phase 
near a magnetic extremum when pulsation amplitude for an oblique pulsator is 
expected to be highest. The methods applied, and the lengths of the data sets 
acquired, would be sufficient for detecting pulsations in all known roAp stars. We 
therefore now reconsider whether HD\,3980 {\it is} a good roAp candidate, when 
comparing its characteristics to those of known roAp stars. If indeed it is a good 
candidate, it either does not pulsate (thus is a noAp star), or we failed to 
detect the pulsations, or, alternatively, the roAp class characteristics are still 
too poorly established to physically discern noAp stars from roAp stars. 

The blue edge of the roAp instability strip where it crosses the main sequence is 
not firmly established. \citet{hubrigetal00} showed that it is around 8500\,K, 
while the theoretical roAp instability strip by \citet{cunha02} extends up to 
around 9500\,K. \citet{ryabchikovaetal04} suggest a transition region (noAp/roAp) 
around 8100\,K where cooler Ap stars may be rapid oscillators. HD\,3980 has a 
precise parallax  ($\pi=14.91\pm0.35$, \citealt{vanleeuwen07}) that for 
\changea{$A_V=0.054$ (NASA/IPAC IRSA maps) gives a luminosity of $\log 
L$/L$_\odot=1.27\pm0.03$ (including all uncertainties). We found an effective 
temperature of $8100\pm200$\,K (estimated error) which places the star well inside 
the theoretical roAp instability strip. 
 The luminosity was derived using 
a bolometric correction, $BC=0.024\pm0.009$, from the relations by \citet{flower96}
for the range of $T_{\rm eff}$ within its error. 
From this luminosity and $T_{\rm  eff}$ we derive a radius of ${\rm R} =
2.19^{+0.19}_{-0.16}$\,R$_{\odot}$. Eq.~5 then gives the magnetic obliquity to be $\beta =
88\fdg7^{+0.8}_{-4.0}$.
Bolometric corrections are notoriously difficult to determine for A stars with
peculiar abundances.
If one similarly uses the bolometric calibration by \citet{landstreetetal07},
accounting for a 0.1\,mag intrinsic error in the calibration, 
the corresponding luminosity and radius become:
$\log L$/L$_\odot=1.25\pm0.07$ and ${\rm R} =
2.14^{+0.30}_{-0.26}$\,R$_{\odot}$, resulting in $\beta = 88\fdg6^{+0.9}_{-28.4}$.
}

\begin{figure} 
\begin{center} 
\epsfxsize 8.5cm\epsfbox{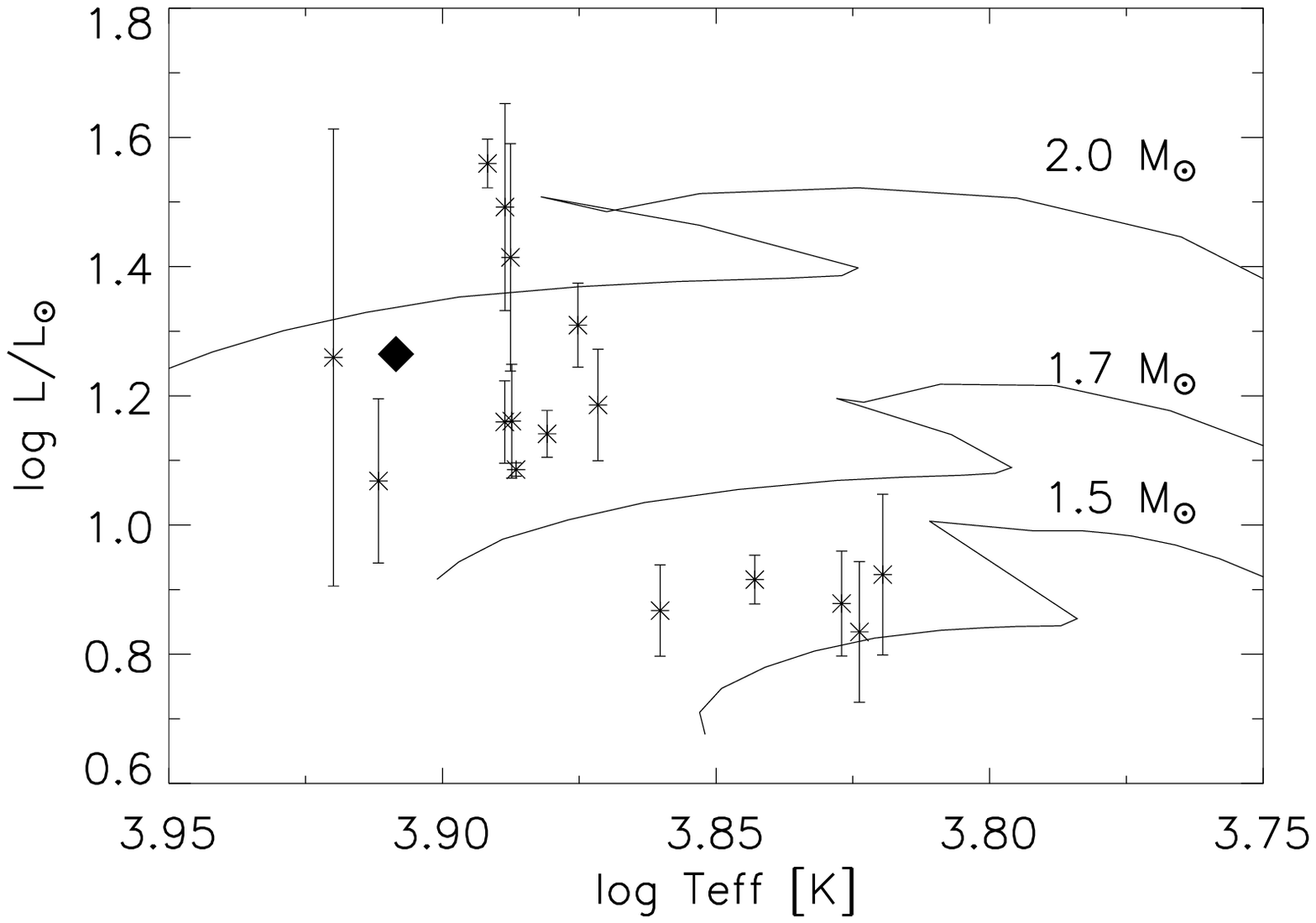} 
\caption{\label{Fig6} A theoretical HR-Diagram for a sample of roAp stars. 
Luminosities were calculated using Hipparcos trigonometric parallaxes 
\citep{perrymanetal97} \changea{except for HD\,3980, for which the revised 
parallax of \citet{vanleeuwen07} was used}. The \citet{moonetal85} calibration of 
Str\"omgren 
photometric indices was employed for estimating the effective temperature. The 
position of HD\,3980 is shown by the filled diamond which is the size of the error 
bar in \changea{$\log L$/L$_\odot$. Solid lines are evolutionary} tracks for stars with 1.5, 1.7 
and 2.0 M$_{\odot}$ taken from \citet{schalleretal92}.} 
\end{center} 
\end{figure} 

The roAp stars have strong magnetic fields that range from several hundred Gauss 
to 24.5\,kG in HD\,154708 \citep{hubrigetal05}. HD\,3980 has a longitudinal field, 
which is within this range. The evolutionary tracks in Fig.~\ref{Fig6}, as well as 
those in figure 1 of \citet{cunha02}, indicate that HD\,3980 in the midst of its 
life on the main sequence, halfway evolved from the ZAMS to the TAMS. 
\citet{cunha02} argued that more evolved roAp stars require stronger magnetic 
fields to suppress the upper envelope convection and enable the rapid oscillations 
to reach observable amplitudes. She further speculated that the magnetic field 
intensities needed for suppressing the convection in roAp stars more often are 
found in roAp stars with magnetically resolved lines. That typically requires a 
magnetic field modulus of $\sim$$3$\,kG \changea{and a slowly rotating Ap star} ($v \sin i=1-
3$\,km\,s$^{-1}$). Based on the longitudinal field measurements of HD\,3980, it 
could have a magnetic field modulus more then 3\,kG. The rotational 
broadening dominates any magnetic splitting of lines at this field intensity and 
direct measurement is not possible. 

\citet{freyhammeretal08} searched for pulsations in 9 evolved cool Ap stars 
located inside the theoretical instability strip, the majority having estimated 
temperatures below $8100$\,K. With similar precision and upper limits on radial 
velocity amplitudes as in this study, these authors found 9 null results. Out of 7 
stars, only 3 had magnetic fields significantly stronger than 2\,kG, which 
possibly explains most of their null results for such evolved stars. The magnetic 
field of HD\,3980 is strong enough to suppress local convection and enable 
pulsational driving of rapid pulsations to reach observable amplitudes. The star 
does not appear to be near the terminal end of its main sequence lifetime, in 
which case this explanation may be less likely. \citet{hubrigetal00} pointed out 
the apparent deficiency of close
binaries among roAp stars (although a `handful' of exceptions are known).
In this context, it is interesting to note that
HD\,3980 is component of a common proper-motion visual binary
(see, e.g.,  \citealt{perrymanetal97}).

Our abundance analysis found that HD\,3980 has strong enhancement of rare earth 
elements, ionisation disequilibria for Nd and Pr and inhomogeneous surface 
distributions of these elements. In these respects, HD\,3980 strongly resembles 
known roAp stars. Only a relatively high temperature, compared to the typical roAp 
stars, may explain why no pulsations are found. We cannot exclude extremely low 
amplitude pulsations of about $5-10$\,m\,s$^{-1}$. Low amplitude oscillations in 
roAp stars are known from, e.g., $\beta$\,CrB \citep{kurtzetal07}, HD\,154708 
\citep{kurtzetal06a} and HD\,116114 \citep{elkinetal05a}. For such low pulsation 
amplitudes, m\,s$^{-1}$ or even sub-m\,s$^{-1}$ high precision measurements are 
required, such as \citet{kochukhovetal07} obtained with HARPS for HD\,75445. This 
in turn requires the stars to be slow rotators, although cross-correlation for 
long spectral regions partly compensates for that. In the case of HD\,3980, even 
the high precision of cross-correlation analysis did not detect low amplitude 
pulsations. 

It is important to find roAp stars with rotation periods of a few days, such as 
for HD\,3980, as they can be subjected to 3D pulsation and abundance studies, such 
as HR\,3831 \citep{kochukhovetal07}. But if the pulsation amplitudes are very low, 
such a study will be limited to very few individual lines or elements and will not 
be very informative. 

We have found that HD\,3980 is a bright, nearby, Li-spotted Ap star with a short 
rotation period which makes it an excellent observing target for \changea{Doppler imaging.}
We demonstrated that the surface distribution is spotted for several 
elements, such as Nd, Eu and Li. We have shown that the star has similar 
characteristics to the known roAp stars, but that it is stable to pulsations with 
an upper limit to the radial velocity amplitudes of individual spectral lines of 
$30$\,m\,s$^{-1}$, less than the amplitudes of the known roAp stars. 

\section{Acknowledgements} 
This research has made use of the NASA/IPAC Infrared Science Archive, which is 
operated by the Jet Propulsion Laboratory, California Institute of Technology, 
under contract with the National Aeronautics and Space Administration 
(http://irsa.ipac.caltech.edu/applications/DUST). This research has made use of 
data obtained using, or software provided by the UK's AstroGrid Virtual 
Observatory Project, which is funded by the Science \& Technology Facilities 
Council and through the EU's Framework 6 programme. LMF, DWK and VGE acknowledge 
support for this work from the Particle Physics and Astronomy Research Council 
(PPARC) and from the Science and Technology Facilities Council (STFC).

{} 

\end{document}